\documentclass[aps,twocolumn,showkeys,amsmath,amssymb,floatfix,pre,superscriptaddress]{revtex4-1}
\usepackage{braket}
\usepackage{graphicx}
\usepackage{booktabs}
\usepackage{multirow}
\usepackage{svg}
\DeclareMathOperator{\Tr}{Tr}
\usepackage{ragged2e}

\usepackage[citecolor=blue,linkcolor=blue,colorlinks=true]{hyperref}

\begin{document}

\title{Double-peak specific heat anomaly and correlations in the Bose-Hubbard model}
\author{Eduardo O. Rizzatti}
\email{eduardo.rizzatti@ufrgs.br}
\affiliation{Instituto de F\'isica, Universidade Federal do 
Rio Grande do Sul, Porto Alegre-RS, Brazil}
\author{Marco Aur\'elio A. Barbosa}
\affiliation{Programa de P\'os-Graduac\~ao em Ci\^encia de Materiais, Universidade de 
 Bras\'ilia, Planaltina-DF, Brazil}
\author{Marcia C. Barbosa}
\affiliation{Instituto de F\'isica, Universidade Federal do 
Rio Grande do Sul, Porto Alegre-RS, Brazil}


\begin{abstract}
\indent \textit{Considering the thermodynamics of bosons in a lattice described by the Bose-Hubbard Hamiltonian, we report the occurrence of anomalous double peaks in their specific heat dependence on temperature. This feature, usually associated with a high geometrical frustration, can also be a consequence of a purely energetic competition.  By employing self-energy functional calculations combined with finite-temperature perturbation theory, we propose a mechanism based on ground-state degeneracies expressed as residual entropies. A general decomposition of the specific heat regarding all possible transitions between the system's eingenvalues provides an insight into the nature of each maximum. Furthermore, we address how the model parameters modify the structure of these peaks based on its spectral properties and atom-atom correlation function.} 
\end{abstract}


\maketitle

\section{Introduction}

\indent In the field of condensed matter physics, the specific heat is a valuable physical observable that provides general information regarding the energy spectrum of a system, a key to its microscopic details~\cite{Gopal1966SpecificTemperatures,Souza2016SpecificModel}. It encodes information on the entropy, a useful thermodynamic quantity directly connected to such microscopic degrees of freedom, in general inaccessible by direct measurements. Among interesting anomalous properties, the appearance of a second peak in the specific heat at low temperatures, known as the Schottky-type anomaly, is experimentally and theoretically observed in several frustrated systems. Typical examples are the magnetic systems on geometrically frustrated lattices with kagome, triangular, or pyrochlore structures. Experimentally, the double-peak anomaly is measured in magnetic pyrochlore oxides~\cite{Gardner2010MagneticOxides},  the canonical spin-ice materials  Dy$_2$Ti$_2$O$_7$~\cite{Ramirez1999Zero-pointIce,Matsuhira2002AField,Hiroi2003SpecificDy2Ti2O7} and Tb$_2$Ti$_2$O$_7$~\cite{Hamaguchi2004Low-temperatureTb2Ti2O7}, their mixtures Dy$_{2-x}$Tb$_x$Ti$_2$O$_7$ ~\cite{Ke2009MagnetothermalDy2-xTbxTi2O7}, lead-based pyrochlores~\cite{Hallas2015MagneticPyrochlores}, and spin glasses like R$_2$Mo$_2$O$_7$ (R$=$Y, Sm, or Gd)~\cite{Raju1992Magnetic-susceptibilityGd}. Other examples include heavy-fermion compounds~\cite{Yamanaka2012HeatField,Lucas2017EntropyCePdAl}, bosonic superfluids in spin-dimer networks~\cite{Brambleby2017AdiabaticBehavior}, CO$_2$N$_2$ plasma~\cite{Zhong2016EffectsPlasma}, lipid bilayers containing colesterol~\cite{Ipsen1989TheoryCholesterol}, as wells as mixtures of liquid crystal and nanoparticles~\cite{Jesenek2011DoubleNanoparticles}. In a theoretical framework, the anomaly was verified for spin models with antiferromagnetic Heisenberg interactions~\cite{Elser1989NuclearSolid,Parkinson2000SmallExchange,Syromyatnikov2004Double-peakClusters,Konstantinidis2005AntiferromagneticSymmetry,Efremov2006SpinMagnets,Hucht2011EffectClusters}, Ising pyrochlore magnets using Monte Carlo simulations~\cite{Harris1998Liquid-GasFerromagnet,Melko2001Long-RangeIce}, Ising models~\cite{Mejdani1996LadderCapacity,Karlova2016ThePolyhedra,Jurcisinova2018HighlySystems,Jurcisinova2018MultipeakAnalysis} in distinct geometries, and quantum ferrimagnets~\cite{Nakanishi2002IntrinsicFerrimagnets}.\\
\indent Since the entropy dependence on temperature determines the specific heat of a given system, this thermodynamic quantity is a relevant ingredient to the presented anomaly. Fundamentally, the geometrical frustration arises from a conflict between the interaction degrees of freedom and the underlying crystal geometry~\cite{Moessner2006GeometricalFrustration}. The described frustration leads to a macroscopic degeneracy computed as a ground-state finite entropy, the so-called residual entropy. Linus Pauling provided one of the first examples of geometrical frustration when describing the low-temperature ordering of protons in water ice I$_h$~\cite{Pauling1935TheArrangement}. The many ways of satisfying the lowest-energy state which reconcile the crystal structure of ice with the known bond lengths were given by the  Bernal-Fowler rules~\cite{Bernal1933AIons}. Based on their prescription, Pauling calculated the ground-state entropy per hydrogen atom of $(1/2)R\ln(3/2) \approx 1.68$ J mol$^{-1}$K$^{-1}$. Interestingly, a similar physical mechanism is verified for the already mentioned highly frustrated pyrochlore magnets, constituting the named spin-ice materials~\cite{Ramirez1999Zero-pointIce,Giauque1936The273K.}. The particular ways these systems can fluctuate between such multiple ground-state configurations can be responsible not only for the anomalies addressed but also for the emergence of all sorts of novel behaviors in fluid and solid phases, including even an artificial edition of electromagnetism~\cite{Balents2010SpinMagnets}. \\
\indent In this work, we report the occurrence of the double-peak specific heat anomaly in the Bose-Hubbard model based on theoretical calculations. The Bose-Hubbard model~\cite{Fisher1989BosonTransition,Jaksch1998a,Krutitsky2016UltracoldLattices} is a paradigm system studied in quantum mechanics. Since its experimental realization in optical lattices~\cite{Greiner2002a}, the ultracold atomic physics has been developing a myriad of experimental and theoretical tools currently used to investigate quantum phase transitions, quantum coherence, and quantum computation~\cite{Jaksch1999EntanglementCollisions,Briegel2000QuantumAtoms,Sachdev2011QuantumTransitions}. Also, a great advantage regarding the arena of optical lattices is the sharp experimental control over parameters such as dimensionality, lattice structure, composition, and atomic interactions~\cite{Windpassinger2013EngineeringLattices,Lewenstein2007UltracoldBeyond,Hofstetter2006UltracoldSystems,Dutta2015Non-standardReview}. More specifically, the high-definition \textit{in situ} imaging acquired through absorption~\cite{Gemelke2009,Zhang2012ObservationLattices} and fluorescence techniques~\cite{Bakr2009ALattice,Sherson2010Single-atom-resolvedInsulator} enables a precise evaluation of the specific heat~\cite{Yang2017QuantumGases}. Here we present an alternative mechanism of frustration observed in this model, devoting some attention to the influence of correlations created by a finite tunneling amplitude. In the previously mentioned systems, the frustration essentially derives from the incompatibilities between geometry and interactions; in the Bose-Hubbard model, the frustration is a consequence of the competition between lattice occupation and local interactions. The main ingredient, a ground-state residual entropy, is also present. Our methodology comprehends numerical tools for treating the many-body thermodynamics of the bosons: self-energy functional theory (SFT) calculations~\cite{Hugel2016BosonicTheory,Rizzatti2020QuantumGases} (closely related to BDMFT approaches) and finite-temperature perturbation theory (PT)~\cite{Metzner1991Linked-clusterModel,Bradlyn2009EffectiveLattices}. \\
\indent The paper is structured as follows. In Sec.~\ref{sec:model} we describe the Bose-Hubbard model, its relevant parameters, and the methodologies employed in our numerical calculations. The double-peak structure of the specific heat is first discussed for the atomic limit in Sec.~\ref{sec:atomic_limit}. The emergence of residual entropies and a specific heat decomposition based on all possible transitions between the energy levels of the spectrum are also discussed. Section~\ref{sec:finite_hopping} addresses the effects of a finite hopping amplitude, including the analysis of spectral and correlation functions. Our main findings and final considerations are summarized in Sec.~\ref{sec:conclusions}.

\section{Model and Methods}\label{sec:model}

\indent In the Bose-Hubbard model~\cite{Fisher1989BosonTransition,Jaksch1998a,Krutitsky2016UltracoldLattices}, the itinerant bosons occupying the lowest-energy band in a lattice are described by the Hamiltonian 
\begin{eqnarray}
H = - J \sum_{\langle i, j \rangle}{b}^{\dagger}_i{b}^{\vphantom{\dagger}}_j + \frac{U}{2} \sum_{i} n_i(n_i-1)- \mu \sum_{i} n_i \;,
\label{bose_hubbard_hamiltonian}
\end{eqnarray}
where $b^{\dagger}_i$, $b_i$, $n_i$ represent the bosonic creation, annihilation, and number operators at site $i$, respectively; $\mu$ is the chemical potential. The parameter $U$ designates the on-site interaction (typically repulsive, taking positive values) and $J$ accounts for the hopping amplitude, a kinetic term involving the probability of tunneling between first neighbor sites. \\
\indent We explore simple cubic~\cite{Greiner2002a} and square lattice~\cite{Gemelke2009,Sherson2010Single-atom-resolvedInsulator} configurations, extensively studied in experiments throughout the last 20 years. The thermodynamic properties of the bosons are mapped through a variational and non-perturbative self-consistent approach, the self-energy functional theory derived by H\"{u}gel \textit{et al}.~\cite{Hugel2016BosonicTheory,Rizzatti2020QuantumGases}, based on the original works for fermions by Potthoff~\cite{Potthoff2003Self-energy-functionalElectrons}. The formalism, which contemplates the $U(1)$ symmetry breaking and BDMFT approaches~\cite{Byczuk2008CorrelatedPhases,Hubener2009MagneticLattice,Hu2009DynamicalModel,Anders2011DynamicalBosons,Snoek2013BosonicTheory}, relies on successive Legendre transformations of the free-energy functional $\Omega$ producing a new functional of the self-energies. In a general panorama, such approximation strategy constricts the self-energy domain to a subspace of self-energies regarding a simpler reference system. Consequently, the original problem is simplified into seeking stationary solutions of this new functional, which is exact, in terms of the reference system's free propagators. When appropriate, we also complement our results with finite-temperature perturbation theory around the atomic limit. Both methods are detailed in the Supplemental Material.\\
\indent Given the equilibrium free energy $\Omega$ obtained from the previously cited techniques, the specific heat is explicitly given by
\begin{eqnarray}
c_{\mu}=-\frac{T}{N_s}\left(\frac{\partial^2 \Omega}{\partial T^2}\right)_{\mu} = T\left(\frac{\partial s}{\partial T}\right)_{\mu} \;,
\label{specific_heat}
\end{eqnarray}
where $T$ is the temperature and $s$ is the thermodynamic entropy per site
\begin{eqnarray}
s= - \frac{1}{N_s}\left(\frac{\partial \Omega}{\partial T}\right)_{\mu} \;, 
\label{entropy}
\end{eqnarray}
considering $N_s$ sites at a fixed chemical potential $\mu$.

\section{The Atomic Limit ($J=0$)}\label{sec:atomic_limit}

\begin{figure}
\centering
\includegraphics[clip,scale=0.9]{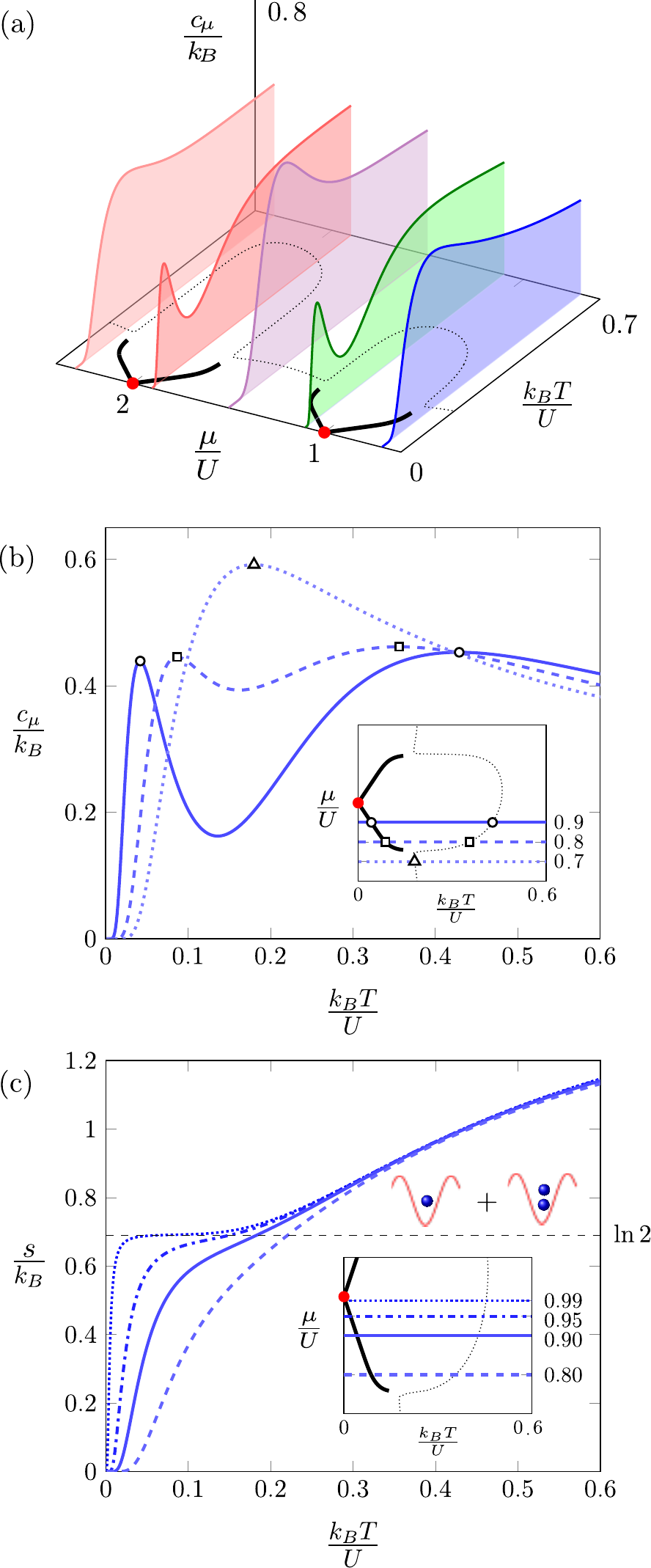}
\caption{Finite-temperature analysis of the specific heat and entropy in the atomic limit $J=0$. (a) The 3D diagram portrays the specific heat $c_\mu$ as a function of the temperature $T$ in a wide range of chemical potential values. The continuous and dotted black curves in the $\mu T$ plane represent the loci of the maximum values attained by $c_{\mu}$. Red dots denote ground-state phase transitions between Mott insulators. In (b) is shown a detailed vision on the temperature dependence of $c_\mu$ considering the values $\mu= 0.7U$, $0.8U$ and $0.9U$, portrayed by the inset. Panel (c) contains the entropy per site $s$ as temperature varies for $\mu= 0.8U$, $0.9U$, $0.95U$ and $0.99U$. }
\label{fig1}
\end{figure}
\begin{figure*}
\centering
\includegraphics[clip,scale=0.85]{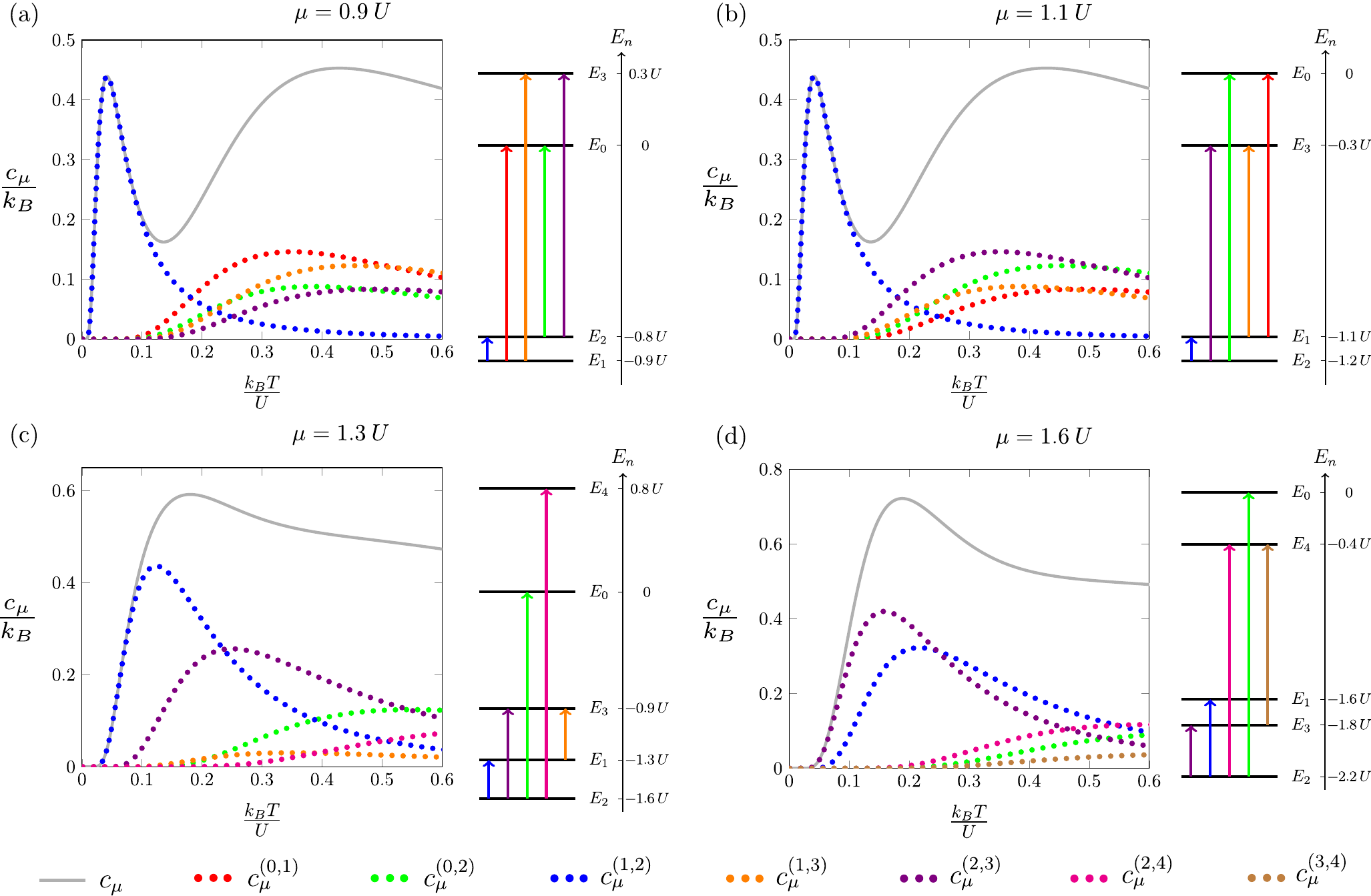}
\caption{The specific heat decomposition and its temperature behavior is exhibited considering four fixed values of chemical potential: (a) $\mu=0.9U$, (b) $\mu=1.1U$, (c) $\mu=1.3U$ and (d) $\mu=1.6U$. The dotted curves represent the contributions $c^{(n,m)}_{\mu}$ to the total specific heat $c_{\mu}$,  which is shown as the continuous curve. Each panel contains a diagram of energy levels assigning a different color to each transition $\ket{n} \rightarrow \ket{m}$ (with energies $E_n \rightarrow E_m$), referring to the respective term $c^{(n,m)}_{\mu}$.  }
\label{fig2}
\end{figure*}
In the following, we show the double-peak structure of $c_\mu$ as a function of $T$, considering the atomic limit ($J=0$). In the absence of the hopping amplitude, there is no superfluid phase and the bosons are found in a normal fluid state. The Hamiltonian described by Eq.~(\ref{bose_hubbard_hamiltonian}) becomes 
\begin{eqnarray}
H^{(0)} = \frac{U}{2} \sum_{i} n_i(n_i-1)- \mu \sum_{i} n_i \;,
\label{bose_hubbard_hamiltonian_atomic_limit}
\end{eqnarray}
a sum of single-site Hamiltonians $H^{(0)}_i$ which can be diagonalized by the number operators eigenvectors $\ket{n_i}$. Thus, the energy eigenvalue of a single site with occupation $n$ is expressed by
\begin{eqnarray}
E_n = \frac{U}{2} n (n -1 ) -\mu n \;.
\label{energy_levels}
\end{eqnarray}
In this scenario, the grand-canonical partition function  
\begin{eqnarray}
\mathcal{Z}^{(0)} (T,\mu) = \Tr\left[e^{-\beta H^{(0)}}\right] = \prod\limits_{i=1}^{N_s}\mathcal{Z}_i^{(0)} = \left[\mathcal{Z}_1^{(0)}\right]^{N_s} 
\label{partition_funciton_atomic_limit}
\end{eqnarray}
becomes simply a product of single-site partition functions
\begin{eqnarray}
\mathcal{Z}^{(0)}_1 (T,\mu) = \sum_{n=0}^\infty  e^{-\beta E_n} \;,
\label{partition_funciton_atomic_limit1}
\end{eqnarray}
where $\beta = 1/k_B T$ and $k_B$ is the Boltzmann constant. The free-energy potential $\Omega^{(0)}$ follows directly from Eqs.~(\ref{partition_funciton_atomic_limit}) and (\ref{partition_funciton_atomic_limit1}) according to
\begin{eqnarray}
\Omega^{(0)} (T, \mu) = -\frac{1}{\beta} \ln \mathcal{Z}^{(0)} =-\frac{N_s}{\beta} \ln \mathcal{Z}^{(0)}_1 \;.
\label{free_energy_atomic_limit}
\end{eqnarray}
\indent Fig.~\hyperref[fig1]{1(a)} portrays the specific heat $c_\mu$ as a function of the temperature $T$ in a three-dimensional diagram including a wide range of chemical potential values (in different colors), with the loci of maxima represented as dotted and continuous black lines in the $\mu T$ plane. At zero temperature, there are ground-state phase transitions (GSPT) between Mott insulators of successive occupation numbers whenever the chemical potential $\mu$ takes on integer values of the local interaction $U$~\cite{Rizzatti2018a,Rizzatti2020QuantumGases}, shown as red dots. At higher temperatures, it is observed a maximum at any value of fixed $\mu$, symbolized by dotted lines. However, as the chemical potential approaches integer values of the interaction, the specific heat develops another peak at low temperatures (the continuous black lines), which is connected to its corresponding GSPT. The low- and high-temperature maxima start to merge together as the chemical potential values move away from the integers values of $U$, finally coalescing in a single peak according to Fig.~\hyperref[fig1]{1(b)}. \\
\indent The entropy dependence on temperature illustrated in Fig.~\hyperref[fig1]{1(c)} suggests that the low-temperature maxima are related to a residual entropy per site of $k_B\ln 2$. This zero-point entropy is due to the degeneracy established at the GSPT between states with different occupations. As the chemical potential approaches the integer multiples of the local interaction, the entropy curves develop a sharp step towards the residual value $k_B\ln 2$ which induces a change in their concavities. This behavior leads to a Schottky-like peak at low temperatures close to the critical points. \\
\indent A deeper understanding on the evolution of the reported double-peak structure can be attained by the following decomposition
\begin{eqnarray}
c_{\mu} = \sum_{n < m} c_{\mu}^{(n,m)} \;,
\label{decomposition_specific_heat}
\end{eqnarray}
demonstrated in Appendix~\ref{appendix:decomposition}. The defined partial specific heat $c_\mu^{(n,m)}$ describes the fluctuations between the energy levels $m$ and $n$ according to
\begin{eqnarray}
c_\mu^{(n,m)} = k_B\frac{\beta^2}{\left[\mathcal{Z}_1^{(0)}\right]^2} (E_n-E_m)^2 \;e^{-\beta (E_n+E_m)}\;.
\label{decomposition_nm}
\end{eqnarray}
Since the specific heat accounts for fluctuations in energy with respect to its mean value, the intuitive idea brought by this decomposition is to visualize the fluctuations in energy as a result of transitions between all possible energy levels. This outcome allows us to understand how these peaks behave and influence each other, isolating the relevance of each transition. \\
\begin{figure*}
\centering
\includegraphics[clip,scale=0.58]{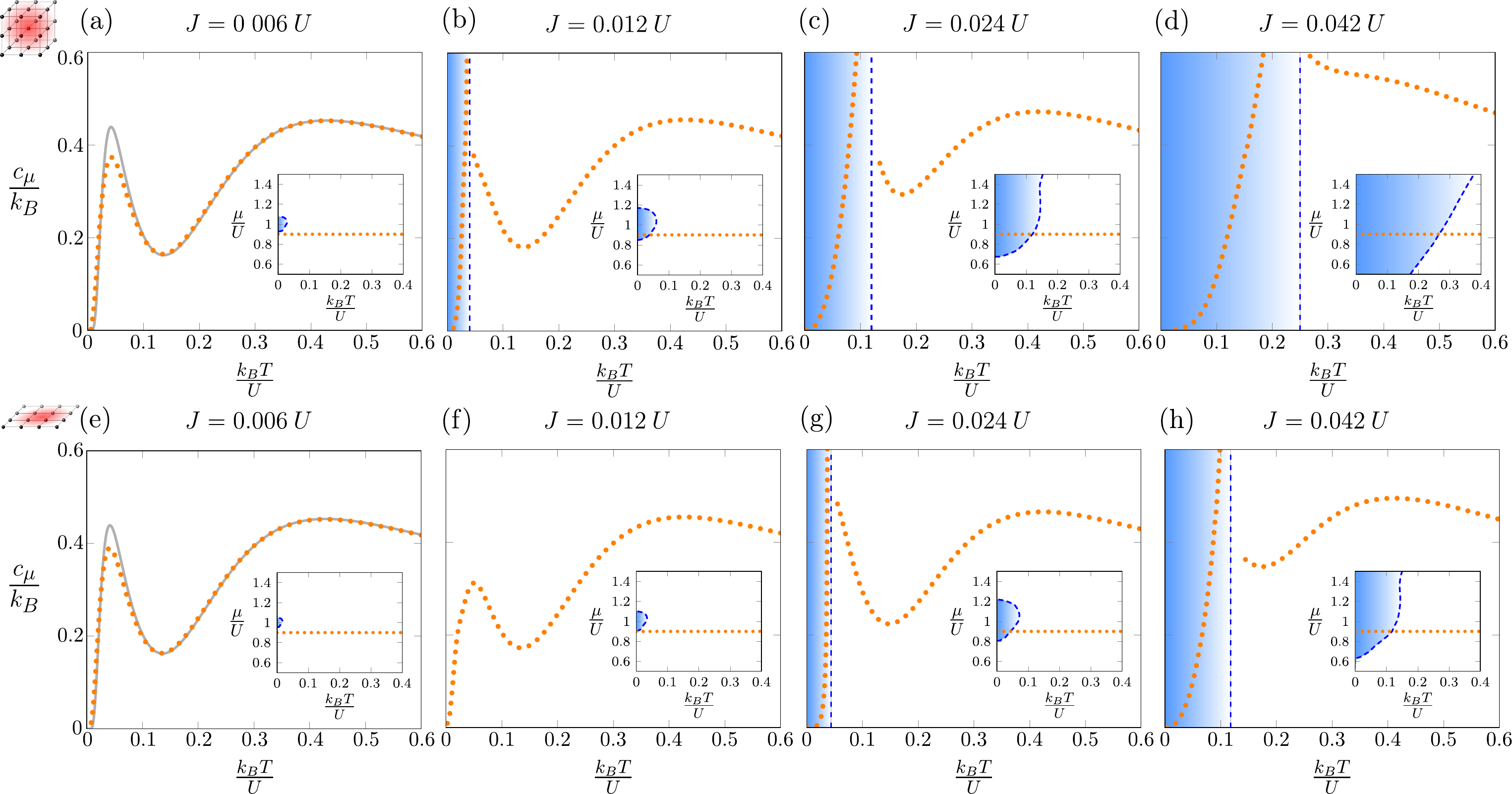}
\caption{The specific heat $c_\mu$ is shown as a function of the temperature $T$ at constant $\mu=0.9U$ for simple cubic (a)-(d) and square lattices (e)-(h) considering four different hopping amplitudes: (a) and (e) $J=0.006U$ (the gray continuous lines are the corresponding atomic limits), (b) and (f)  $J=0.012U$, (c) and (g)  $J=0.024U$, (d) and (h)  $J=0.042U$. The superfluid domain is represented by the blue area, with the phase boundaries symbolized by the blue dashed lines. The insets exhibit the respective $\mu T$ phase diagrams.}
\label{fig3}
\end{figure*}
\indent In Fig.~\ref{fig2} we present the specific heat as a function of temperature by continuous lines including four different values of chemical potential: (a) $\mu=0.9U$, (b) $\mu=1.1U$, (c) $\mu=1.3U$ and (d) $\mu=1.6U$. For each scenario, the most relevant partial contributions $c_{\mu}^{(n,m)}$ are addressed as the colored dotted lines, with the respective level transitions depicted accordingly. Figure~\hyperref[fig2]{2(a)} shows the $\mu=0.9U$ case. The low-temperature peak appears due to the $\ket{1} \rightarrow \ket{2}$ transition. Other contributions add up to form the second maximum at a higher temperature; among them, the most relevant terms in this temperature range arise from the transitions $\ket{1} \rightarrow \ket{0}$ (red) and $\ket{1} \rightarrow \ket{3}$ (orange) which connect the states $\ket{0}$ and $\ket{3}$ to the ground state $\ket{1}$, respectively. In Fig.~\hyperref[fig2]{2(b)}, for $\mu=1.1U$, the ket $\ket{2}$ becomes the new ground sate and the first peak persists due to the available transition $\ket{2} \rightarrow \ket{1}$. As a result of this new ground state, the contributions for the second maximum from $\ket{2} \rightarrow \ket{0}$ (green) and $\ket{2} \rightarrow \ket{3}$ (purple) become more prominent than the previous transitions $\ket{1} \rightarrow \ket{0}$ (red) and $\ket{1} \rightarrow \ket{3}$ (orange). In Fig.~\hyperref[fig2]{2(c)}, where $\mu=1.3U$, the transition $\ket{2} \rightarrow \ket{3}$ (purple) starts to gain relevance since the energy level $E_3$ approaches $E_1$. As a consequence of these two competing contributions, the specific heat develops only one maximum. Figure~\hyperref[fig2]{2(d)}, which displays $\mu=1.6U$, finally shows $E_3$ becoming the first excited state, with $c^{(2,3)}$ the most relevant term at low temperatures. By increasing the chemical potential, another analogous cycle starts presenting the same fundamental mechanisms already explored. 

\section{Finite Hopping ($J\neq 0$)}\label{sec:finite_hopping}

\indent Including finite-hopping effects, the specific heat behavior with temperature for simple cubic and square lattices is illustrated in Figs.~\hyperref[fig3]{3(a)}-\hyperref[fig3]{(d)} and Figs.~\hyperref[fig3]{3(e)}-\hyperref[fig3]{(h)}, respectively. The observables were evaluated using the previously described SFT formalism. In this framework, we considered four values of tunneling: $J=0.006U$, $J=0.012U$, $J=0.024U$ and $J=0.042U$, at a fixed chemical potential $\mu=0.9U$.  Physically, such increasing tunneling amplitudes correspond to decreasing lattice potential depths, going from very deep lattices to shallower ones. The introduction of the kinetic term $J$ gives rise to a superfluid phase which grows from the GSPT points~\cite{Freericks1994PhaseModel}. The superfluid phase is represented in the blue area, with the superfluid to normal phase boundaries shown as dashed blue lines. The respective insets portray the $\mu T$ phase diagrams.  \\  
\indent First we analyze the three-dimensional case, exhibited in Figs.~\hyperref[fig3]{3(a)}-\hyperref[fig3]{(d)}. Considering the chosen value $J=0.006U$ in Fig.~\hyperref[fig3]{3(a)}, the system is found in the normal phase and the two-peak structure of the specific heat remains present. For comparison reasons, we also indicate the corresponding atomic limit result as the gray continuous line. It is observed a clear reduction of the first peak when the hopping is slightly turned on; the second one, however, does not present relevant quantitative differences. Intuitively, as the low-temperature maximum derives from the energetic competition between two states, the introduction of the hopping breaks this degeneracy, mitigating its magnitude. A detailed analysis regarding such behavior is also developed in Subsecs.~\ref{subsec:spectral_functions} and~\ref{subsec:correlations}, involving the analysis of spectral properties and correlation functions. By increasing the hopping to $J=0.012U$, Fig.~\hyperref[fig3]{3(b)} shows the appearance of the superfluid phase. Inasmuch as response functions tend to grow in magnitude near continuous phase transitions, we detect an increasing of $c_{\mu}$ in the neighborhood of the phase transition. Consequently, the first peak turns into a critical divergence while the second one is still observed.  Figure~\hyperref[fig3]{3(c)} portrays the $J=0.024U$ scenario, where $c_\mu$ increases monotonically in the superfluid and the second maximum remains to be seen in the normal phase. For a larger hopping term $J=0.042U$ in  Fig.~\hyperref[fig3]{3(d)}, the superfluid domain expands to an extent that the high-temperature peak disappears. Thus the two original maxima, born in the the atomic limit, are only verified in a perturbative regime. \\
\indent Considering the same hopping values and $\mu=0.9U$,  Figs.~\hyperref[fig3]{3(e)}-\hyperref[fig3]{(h)} depict the specific heat dependence on temperature for the square lattice configuration. The superfluid domain for each tunneling amplitude is smaller when compared to the three-dimensional picture. Indeed, since there is a reduction on the number of available directions for a particle to hop in a square lattice, the superfluid region shrinks. Therefore, it takes larger values of $J$ to observe in a square lattice the same effects seen in a simple cubic one. For $J=0.006U$, Fig.~\hyperref[fig3]{3(e)} shows two peaks in the normal phase, with the low-temperature one appearing reduced in comparison to the atomic limit (continuous line). Increasing $J$, Fig.~\hyperref[fig3]{3(f)} shows that such peak is diminished even further for $J=0.012U$. Finally, Figs.~\hyperref[fig3]{3(g)} and~\hyperref[fig3]{3(h)} demonstrate that the rise of the superfluid phase starts to destroy the developed peaks. \\
\indent In the next Subsections we analyze the two peaks of $c_{\mu}$ in the presence of hopping in greater detail. The $J=0.006U$ case, which preserves such property in two and three dimensions, is explored inside the normal phase.

\begin{figure}
\centering
\includegraphics[clip,scale=1]{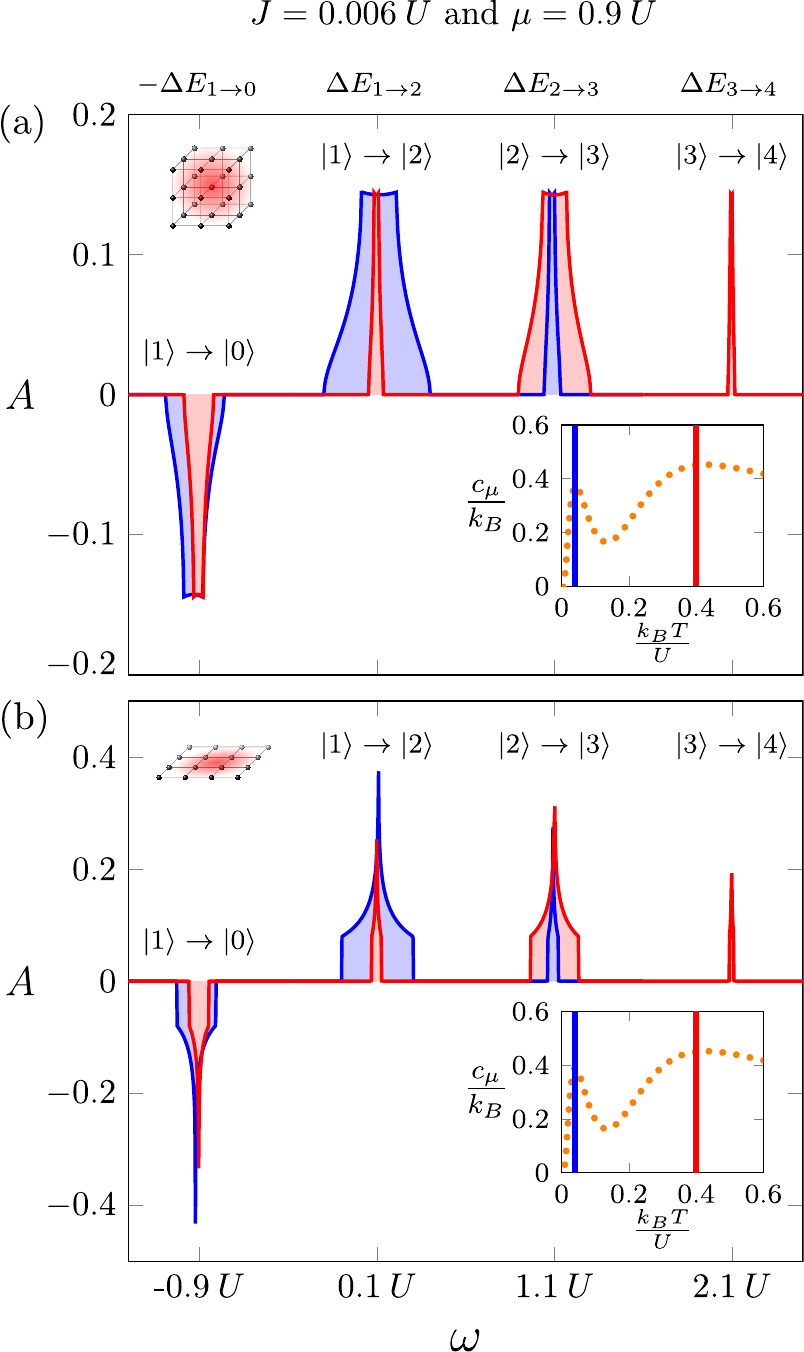}
\caption{The local spectral function $A$ versus the real frequency $\omega$ at fixed $J=0.006U$ and $\mu=0.9U$, considering simple cubic (a) and square lattices (b). The two temperatures addressed $k_BT=0.04U$ (blue) and $k_BT=0.4U$ (red), located around each maximum of the specific heat, are marked in the inset panel. Above each resonance, the most significant transitions $\ket{n} \rightarrow \ket{m}$ are identified.}
\label{fig4}
\end{figure}

\subsection{Spectral Functions}\label{subsec:spectral_functions}

\indent As spectral functions reveal how states occupy a given energy interval, they also provide useful information on the specific heat behavior. The spectral function $A(\mathbf{k},\omega)$, a generalized density of states in frequency $\omega$ and momentum space $\mathbf{k}$, is defined by the imaginary part of the retarded Green's function~\cite{Stefanucci2013NonequilibriumSystems} according to 
\begin{eqnarray}
A(\mathbf{k},\omega) = - \frac{1}{\pi} \Im[G^{R}(\mathbf{k},\omega)] \;.
\label{spectral_function}
\end{eqnarray}
This follows from the Matsubara Green's function by analytic continuation, applying the prescription $i\omega_n \rightarrow \omega + i0^{+}$ where $\omega_n=2\pi n/\beta$. Its integration over the momentum yields the local spectral function
\begin{eqnarray}
A(\omega) = \frac{1}{N_s} \sum_{\mathbf{k}} A(\mathbf{k},\omega) \;.
\label{local_spectral_function}
\end{eqnarray}
Concerning other range of parameters, the spectral properties of the Bose-Hubbard were also discussed using SFT~\cite{Hugel2016BosonicTheory}, as well as employing QMC and BDMFT methods ~\cite{Panas2015NumericalTheory,Strand2015BeyondBosons}. In Figs.~\hyperref[fig4]{4(a)} and~\hyperref[fig4]{4(b)}, we show the behavior of $A$ as $\omega$ is varied considering $\mu=0.9U$ and $J=0.006U$ in the SFT framework for simple cubic and square lattices, respectively. These are the same scenarios reported by Figs.~\hyperref[fig3]{3(a)} and~\hyperref[fig3]{3(e)}. The temperatures addressed in our analysis, $k_BT = 0.04U$ (blue) and $k_BT = 0.4U$ (red), are located near each maximum of $c_{\mu}$ as the inset panel illustrates. The noticed resonances can be thought in terms of transitions between local occupation number states, which are eigenvectors of the zero-hopping Hamiltonian explored in Sec.~\ref{sec:atomic_limit}. As demonstrated in Appendix~\ref{appendix:spectral}, the local spectral function in the atomic limit reads as 
\begin{eqnarray}
A^{(0)}(\omega) &=& \frac{1}{\mathcal{Z}^{(0)}} \sum_n \delta (\omega -\Delta E_{n\rightarrow n+1}) (n+1) e^{-\beta E_n} \nonumber \\
&-&\frac{1}{\mathcal{Z}^{(0)}} \sum_n  \delta (\omega -\Delta E_{n-1\rightarrow n})\:n e^{-\beta E_n} \;,
\label{aomic_limit_local_spectral_function}
\end{eqnarray}
a collection of delta functions centered around the energy level transitions $\Delta E_{n\rightarrow m} = E_m-E_n$. With the perturbative inclusion of hopping, the relevant Hubbard bands emerge from such values, with a particular shape reflecting the dimensionality of each case. \\
\indent Regarding the low-temperature regime (blue curve), there are  contributions from only three transitions: singlon-holon $\ket{1}\rightarrow\ket{0}$, singlon-doublon $\ket{1}\rightarrow\ket{2}$ and doublon-triplon $\ket{2}\rightarrow\ket{3}$. Given its larger width, the most relevant of them is the portion $\ket{1}\rightarrow\ket{2}$, because the chosen $\mu$ value is close to the transition between states of occupation number $1$ and $2$. These states at low temperatures produce the Schottky peak as explained in the $J=0$ scenario. Regarding the second maximum (red curve), other states of higher energy  become available such as the transition triplon-quadruplon $\ket{3}\rightarrow\ket{4}$, while the transition singlon-doublon $\ket{1}\rightarrow\ket{2}$ are less probable to happen. 

\subsection{Correlations and the Double Peaks}\label{subsec:correlations}

\begin{figure}
\centering
\includegraphics[clip,scale=0.95]{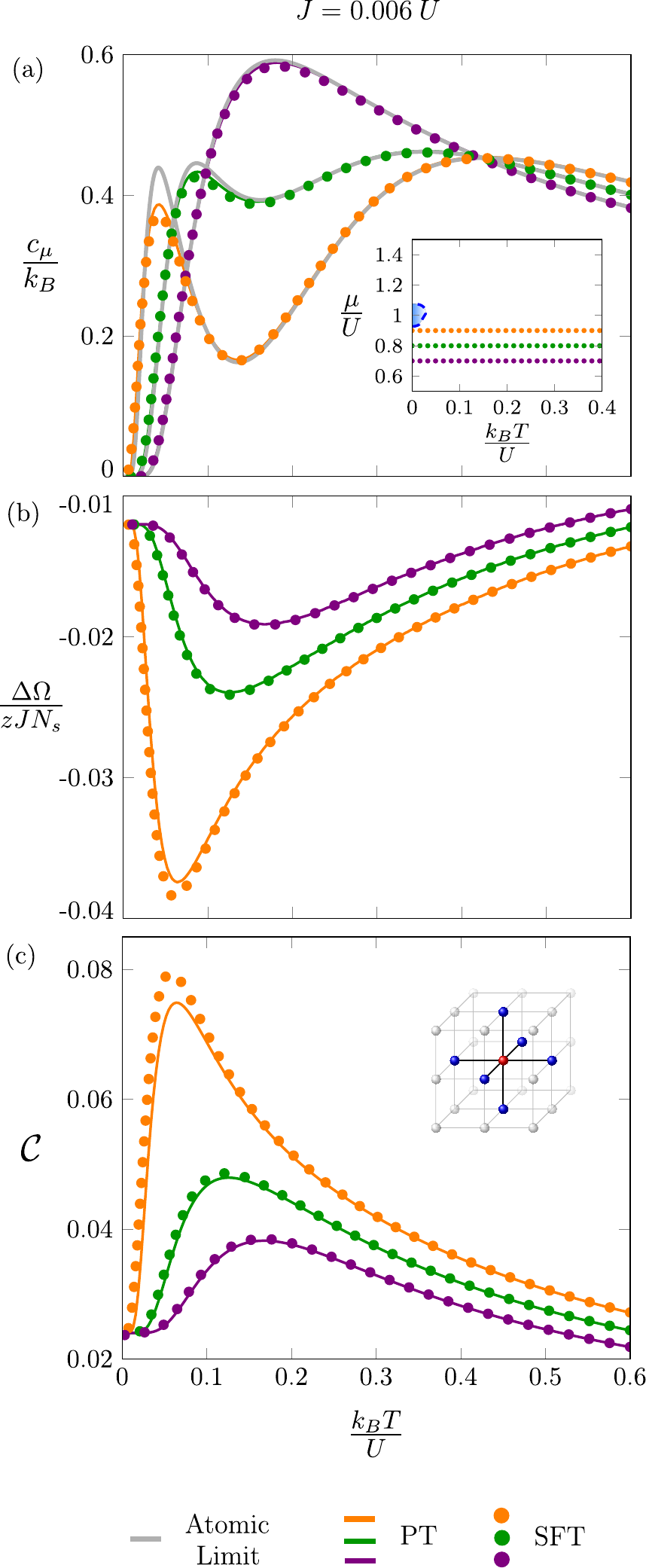}
\caption{The specific heat $c_\mu$ (a), free energy variation $\Delta \Omega$ (b) and atom-atom correlation $\mathcal{C}$ (c) are shown as functions of temperature at $J=0.006U$. As the inset phase diagram $\mu T$ reveals, three values of chemical potential are considered: $\mu=0.7U$ (purple), $\mu=0.8U$ (green), and $\mu=0.9U$ (orange). The dots represent SFT calculations while the respective colored continuous lines symbolize the PT results. The atomic limit of $c_\mu$ is shown as gray continuous curves in (a).}
\label{fig5}
\end{figure}
\begin{figure*}
\centering
\includegraphics[clip,scale=0.9]{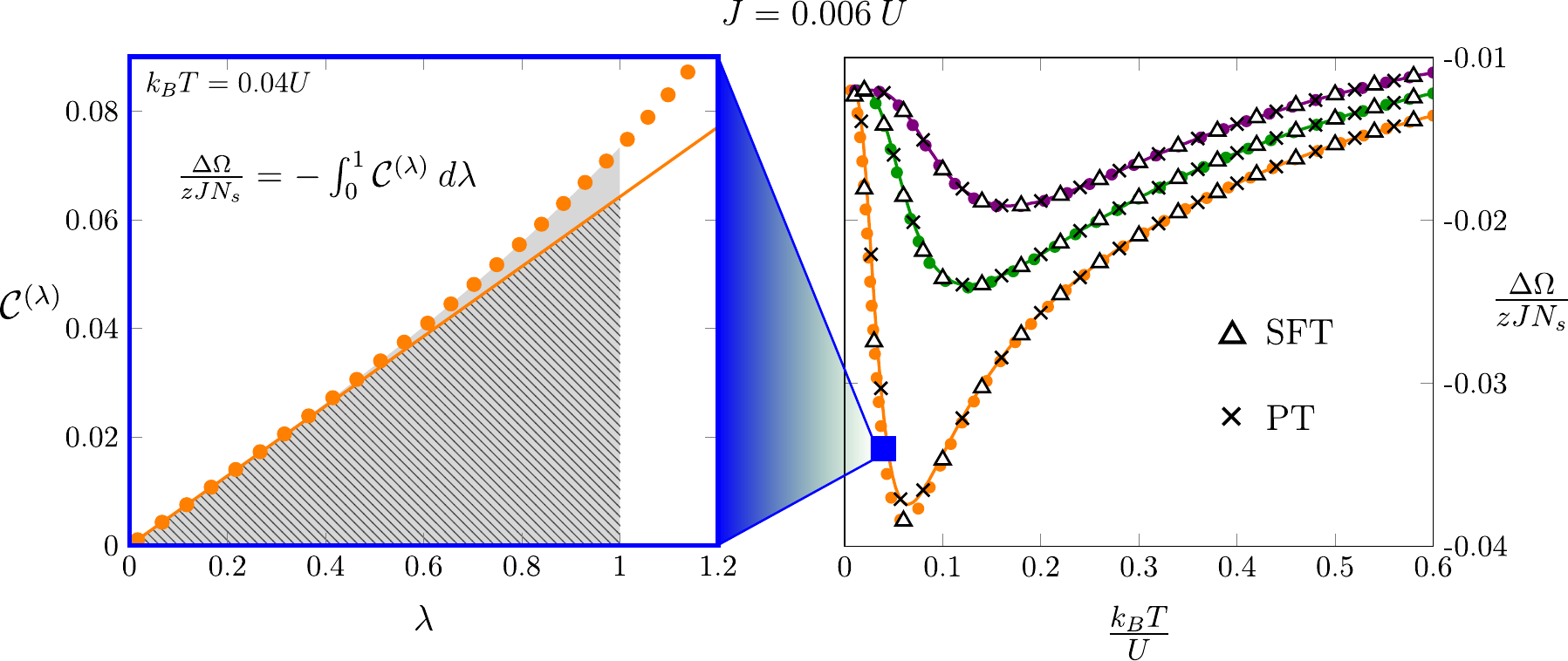}
\caption{In the right panel, $\Delta \Omega$ is plotted against temperature $T$ showing the agreement between points (crosses and triangles) evaluated through Eq.~(\ref{delta_omega_correlation}) and the curves from Fig.~\hyperref[fig5]{5(b)}. Crosses and triangles correspond to PT and SFT calculations, respectively. For $k_BT=0.04U$ and $\mu=0.9U$, the left panel illustrates how these points are determined by integration of the correlation function $\mathcal{C}^{(\lambda)}$.}
\label{fig6}
\end{figure*}

\indent As pointed out by Figs.~\hyperref[fig3]{3(a)} and~\hyperref[fig3]{3(e)}, the first maximum appearing in the specific heat versus temperature plot presents a reduction when compared to the atomic limit. In Fig.~\ref{fig5}, we supply a detailed view on this subject considering the simple cubic setup for deep lattices $J=0.006U$ (the results in the square lattice are quantitatively similar) and three values of chemical potential: $\mu=0.7 U$ (purple), $\mu=0.8 U$ (green), and $\mu=0.9 U$ (orange). Colored dots symbolize SFT calculations while the respective continuous lines represent the perturbation theory until second order on $J/U$. Figure~\hyperref[fig5]{5(a)} exhibits $c_{\mu}$ as $T$ varies in the described scenario, including the zero-hopping cases as continuous gray lines. We observe that the deviations from the $J=0$ situation become more prominent as the chemical potential approaches $\mu=1.0U$. Since the specific heat is essentially a second derivative of the free energy with respect to the temperature, Fig.~\hyperref[fig5]{5(b)} shows the difference $\Delta \Omega=\Omega-\Omega^{(0)} $ of the system's free energy $\Omega$ in the presence of hopping relative to its atomic limit $\Omega^{(0)}$.  When $\mu=0.7U$, the free energy is slightly reduced by the introduction of the tunneling term. However, for $\mu=0.8U$ and $\mu=0.9U$, important variations in a low-temperature regime are verified. The prominent curvature $\Delta \Omega$ implies a relevant second derivative. This information gets translated into a contribution to $c_{\mu}$, responsible for the magnitude reduction in the first maximum.\\
\indent The question remaining is what physical mechanism generates such curvature in the free energy. Since the inclusion of the tunneling term induces correlations among particles located at different sites $j'$ and $j$, we analyze the atom-atom correlation function~\cite{Freericks2009Strong-couplingResults,Teichmann2009Process-chainDiagram}
\begin{eqnarray}
\mathcal{C}_{j'j} = \langle b^{\dagger}_{j'} b_{j^{\phantom{'}}}\vphantom{\dagger} \hspace{-0.1cm}\rangle\;,
\label{correlation_atom_atom}
\end{eqnarray}
related to the imaginary-time Green's function
\begin{eqnarray}
\mathcal{G}_{jj'} (\tau,\tau') = -\langle \mathcal{T}[b_{j^{\phantom{'}}}\vphantom{\dagger}\hspace{-0.1cm}(\tau) b_{j'}^{\dagger}(\tau')] \rangle
\label{causal_greens_function}
\end{eqnarray}
through
\begin{eqnarray}
\mathcal{C}_{j'j} = -G_{jj'}(0,0^{+}) \;.
\label{correlation_greens_function}
\end{eqnarray}
Accordingly, these correlations can be indirectly probed as they are connected to the Fourier transform of the momentum distribution function
\begin{eqnarray}
n(\mathbf{k}) = \frac{|w(\mathbf{k})|^2}{N_s}\sum_{j,j'} \mathcal{C}_{j'j\vphantom{'}} e^{-i\mathbf{k}\dot (\mathbf{r}_{j'}-\mathbf{r}_{j\vphantom{'}})} \;,
\label{momentum_distribution_function}
\end{eqnarray}
where $w(\mathbf{k})$ is the Fourier transform of the Wannier function. The quantity $n(\mathbf{k})$ is measured by the time-of-flight absorption experiments, in which the trapping field is turned off enabling the cloud of atoms to expand during a certain amount of time~\cite{Gerbier2005PhaseInsulator, Hoffmann2009VisibilityTemperatures}. More specifically, we focus our attention on the atom-atom correlation function between first neighbors (since the particles of the model are able to hop to first-neighbor sites), denoted simply as $\mathcal{C}$ due to the system's homogeneity. In Fig.~\hyperref[fig5]{5(c)} we show $\mathcal{C}$ as a function of the temperature, mirroring the parameters of Figs.~\hyperref[fig5]{5(a)} and~\hyperref[fig5]{5(b)}. Similarly to the free energy corrections  $\Delta \Omega$, the correlations become important as we increase the chemical potential towards the integer values of $U$, developing a sharp peak at low temperatures $k_BT \approx 0.1U$.    \\
\indent The impact of the correlations on the free energy can be examined in an analytical perspective. Indeed, it is possible to exactly determine the free energy from the atom-atom correlation function based on the introduction of a continuous coupling parameter, following Refs.~\cite{L.Fetter1971QuantumSystems,Negele1998QuantumSystems}. First, we parametrize the Hamiltonian of Eq.~(\ref{bose_hubbard_hamiltonian}) in the form
\begin{eqnarray}
H^{(\lambda)} = H^{(0)} +\lambda \mathcal{H} \;,
\label{hamiltonian_parameter}
\end{eqnarray}
where $H^{(0)}$ corresponds to the atomic limit defined in Eq.~(\ref{bose_hubbard_hamiltonian_atomic_limit}) and $\mathcal{H}$ is the kinetic term
\begin{eqnarray}
\mathcal{H} =  - J \sum_{\langle i,j \rangle} {b}^{\dagger}_i{b}^{\vphantom{\dagger}}_j \;.
\label{hamiltonian_parameter_w}
\end{eqnarray}
Note that the value $\lambda=0$ correspond to the atomic limit, while $\lambda=1$ recovers the original model considered, with $H^{(1)} = H$. As demonstrated in Appendix~\ref{appendix:correlations_free_energy}, the change in free energy caused by the hopping reads as
\begin{eqnarray}
\Delta \Omega = \int_{0}^{1} \langle \mathcal{H} \rangle_{\lambda} \: d\lambda \;,
\label{delta_omega_v}
\end{eqnarray}
considering thermal averages $\langle \dots \rangle_{\lambda}$ taken with respect to the Hamiltonian of Eq.~(\ref{hamiltonian_parameter}). The definitions from Eqs.~(\ref{hamiltonian_parameter_w}) and (\ref{correlation_atom_atom}) substituted into Eq.~(\ref{delta_omega_v}) lead to the exact expression
\begin{eqnarray}
\Delta \Omega = -zJN_s\int_{0}^{1} \mathcal{C}^{(\lambda)} \: d\lambda \;.
\label{delta_omega_correlation}
\end{eqnarray}
The term $\mathcal{C}^{(\lambda)}=\langle b^{\dagger}_{j'} b_{j^{\phantom{'}}}\vphantom{\dagger}\hspace{-0.1cm} \rangle_{\lambda}$ is simply the atom-atom correlation function between first neighbors $j$ and $j'$, thermally averaged at a hopping value $\lambda J$. This expression associates the theoretical free energy to the atom-atom correlation, an indirect measurable quantity, by a process of charging up the tunneling amplitude until the desired value. Through Fig.~\ref{fig6}, we verify the agreement between both sides of Eq.~(\ref{delta_omega_correlation}) considering SFT and perturbative methods. In particular, perturbation theory up to second order yields the direct relation 
\begin{eqnarray}
\Delta \Omega = -\frac{zJN_s}{2} \:\mathcal{C}+ \mathcal{O}(J^4) \;.
\label{delta_omega_correlation_perturbation}
\end{eqnarray}
Therefore, the low-temperature correlations, which grow near integers multiples of the interaction, cause an important curvature change in the free energy, whose impact is revealed in the reduction of the first maximum of the specific heat.

\section{Conclusions}\label{sec:conclusions}

\indent We studied in detail the specific heat dependence on temperature of bosons described by the Bose-Hubbard model, considering simple cubic and square lattice geometries. Our theoretical analysis comprehended numerical methods, including the self-energy functional theory (a non-perturbative and self-consistent approach, suitable for the description of both normal and superfluid phases) complemented by finite-temperature perturbation theory around the atomic limit. Our results revealed an anomalous double-peak behavior of the specific heat capacity as temperature is varied, connected to a residual entropy established in the atomic limit. Indeed, for $J=0$, the anomaly is present for values of $\mu$ near ground-state phase transitions between Mott insulators of successive occupation number. In this regime, such ground-state macroscopic degeneracy is originated from the energetic competition between the local interaction $U$ and the chemical potential $\mu$. While the local interaction is repulsive and tends to decrease the occupation of a single site, the chemical potential acts in the opposite direction, favoring occupation. Specifically, when $\mu$ takes on integer values of the on-site repulsion (which is exactly the energy required to add another particle at a given site), the system becomes frustrated energetically. This energetic frustration gets translated into an additional low-temperature maximum of $c_{\mu}$ as a function of $T$. In this framework, we also demonstrated a general decomposition of the specific heat, based on all possible transitions realizable between the energy eigenvalues of the system's spectrum. This result enabled us to identify the relevant energy level transitions and how the noted peaks evolve as the chemical potential is varied. \\
\indent For finite hopping amplitudes, the superfluid phase tends to destroy the reported maxima. Indeed, the double-peak structure is only observed inside the normal phase and in a perturbative regime, corresponding to very deep optical lattices.  Additionally, the low-temperature maximum presents a reduced magnitude when compared to zero-hopping scenario. This can be explained by the effects of correlations created between first neighbors (enabled by the hopping), producing a prominent change in the free energy's curvature.  Such connection was established by an exact and general result relating the free energy change to the building up of correlations as the tunneling amplitude is turned on. Spectral functions were also used to illustrate the relevant excitations dominant at each maximum.

\section{Acknowledgments} 

\indent We acknowledge the financial support of the Brazilian funding agencies FAPDF, CNPq and CAPES. 

\appendix

\section{Decomposition of the Specific Heat}\label{appendix:decomposition}

\indent In order to demonstrate Eq.~(\ref{decomposition_nm}), we explore the heat capacity $C$, its extensive related function. In general terms, the heat capacity accounts for fluctuations in energy with respect to its mean value according to 
\begin{eqnarray}
C =-T\left(\frac{\partial^2 \Omega}{\partial T^2}\right) =k_B \beta^2\left ( \langle E^2 \rangle - \langle E \rangle^2\right)\;.
\label{appendix_decomposition_heat_capacity}
\end{eqnarray}
The fluctuations around the mean value $(\delta E)^2 = \langle E^2 \rangle - \langle E \rangle^2$ are explicitly determined by the averages
\begin{eqnarray}
\langle E^2 \rangle = \frac{1}{\mathcal{Z}} \sum_{n} E_n^2 \;e^{-\beta E_n}
\label{appendix_decomposition_average1}
\end{eqnarray}
and
\begin{eqnarray}
\langle E \rangle^2 = \frac{1}{\mathcal{Z}^2} \sum_{n,m} E_n E_m \;e^{-\beta (E_n+E_m)}\;,
\label{appendix_decomposition_average2}
\end{eqnarray}
where 
\begin{eqnarray}
\mathcal{Z} = \sum_{n} \;e^{-\beta E_n}
\label{appendix_decomposition_partition_function}
\end{eqnarray}
is the partition function. Alternatively, it is possible to express Eq.~(\ref{appendix_decomposition_average1}) in the form
\begin{eqnarray}
\langle E^2 \rangle &=& \left ( \frac{1}{\mathcal{Z}} \sum_{n} E_n^2 \;e^{-\beta E_n}\right) \times \left ( \frac{1}{\mathcal{Z}} \sum_{m} \;e^{-\beta E_m}\right ) \nonumber \\
&=& \frac{1}{\mathcal{Z}^2} \sum_{n,m} E_n^2 \;e^{-\beta (E_n+E_m)}\;,
\label{appendix_decomposition_average1_new}
\end{eqnarray}
which allows us to write the heat capacity according to
\begin{eqnarray}
C = k_B\frac{\beta^2}{\mathcal{Z}^2} \sum_{n,m} (E_n^2-E_nE_m) \;e^{-\beta (E_n+E_m)}\;.
\label{appendix_decomposition_heat_capacity_new}
\end{eqnarray}
Since the terms $n=m$ are zero, we separate the sum into two parts 
\begin{eqnarray}
C &=& k_B\frac{\beta^2}{\mathcal{Z}^2} \sum_{n < m} (E_n^2-E_nE_m) \;e^{-\beta (E_n+E_m)} \nonumber \\
&& + k_B\frac{\beta^2}{\mathcal{Z}^2} \sum_{n > m} (E_n^2-E_nE_m) \;e^{-\beta (E_n+E_m)}\;.
\label{appendix_decomposition_heat_capacity_separation}
\end{eqnarray}
By exchanging the labels $n \rightarrow m$ and $m \rightarrow n$ in the second term, we obtain
\begin{eqnarray}
C &=& k_B\frac{\beta^2}{\mathcal{Z}^2} \sum_{n < m} (E_n^2-E_nE_m) \;e^{-\beta (E_n+E_m)} \nonumber \\
&& + k_B\frac{\beta^2}{\mathcal{Z}^2} \sum_{m > n} (E_m^2-E_mE_n) \;e^{-\beta (E_m+E_n)} \nonumber \\
&=& k_B\frac{\beta^2}{\mathcal{Z}^2} \sum_{n < m} (E_n^2-E_nE_m + E_m^2-E_mE_n) \;e^{-\beta (E_n+E_m)} \nonumber \\
&=& k_B\frac{\beta^2}{\mathcal{Z}^2} \sum_{n < m} (E_n-E_m)^2 \;e^{-\beta (E_n+E_m)} \;.
\label{appendix_decomposition_heat_capacity_decomposition}
\end{eqnarray}
Such decomposition impels the definition of the partial specific heat $c^{(n,m)}$, describing the fluctuations between the energy levels $m$ and $n$
\begin{eqnarray}
C^{(n,m)} = k_B\frac{\beta^2}{\mathcal{Z}^2} (E_n-E_m)^2 \;e^{-\beta (E_n+E_m)}\;.
\label{appendix_decomposition_partial_heat_capacity}
\end{eqnarray}
Therefore, the heat capacity becomes simply the sum of the defined partial components
\begin{eqnarray}
C = \sum_{n < m} C^{(n,m)}.
\label{appendix_decomposition_heat_capacity_sum}
\end{eqnarray}
This result enables us to analyze and filter the relevance of each energy level transitions.

\section{Spectral Function in the Atomic Limit}\label{appendix:spectral}

For $J=0$, the local one-particle Green's function in Matsubara space is given by
\begin{eqnarray}
G^{(0)}(i\omega_n) \frac{1}{\mathcal{Z}^{(0)}} \sum_{n,m} \frac{\bra{n} b\ket{m} \bra{m}b^{\dagger}\ket{n}}{i\omega_n +E_n-E_m} ( e^{-\beta E_n}-e^{-\beta E_m} ) \;, \nonumber
\label{appendix_green_function}
\end{eqnarray}
when expressed using the Lehmann representation~\cite{Stefanucci2013NonequilibriumSystems}. For interpretation purposes, it is decomposed into particle and hole excitation branches~\cite{Strand2015BeyondBosons} according to
\begin{eqnarray}
G^{(0)}(i\omega_n)= G^{(0)}_p(i\omega_n)+G^{(0)}_h(i\omega_n) \;,
\label{appendix_green_function_decomposition}
\end{eqnarray}
where
\begin{eqnarray}
G^{(0)}_p(i\omega_n) &=& \frac{1}{\mathcal{Z}^{(0)}} \sum_{n,m} \frac{\bra{n} b\ket{m} \bra{m}b^{\dagger}\ket{n}}{i\omega_n +E_n-E_m} e^{-\beta E_n} 
\label{appendix_green_function_particle}
\end{eqnarray}
and
\begin{eqnarray}
G^{(0)}_h(i\omega_n) &=& \frac{1}{\mathcal{Z}^{(0)}} \sum_{n,m} \frac{\bra{m} b\ket{n} \bra{n}b^{\dagger}\ket{m}}{i\omega_n +E_m-E_n} e^{-\beta E_n} .
\label{appendix_green_function_hole}
\end{eqnarray}
By analytical continuation, we determine the retarded Green's function $G^{(0)}_R (\omega) = G^{(0)}(\omega +i\eta)$, with $\eta\rightarrow 0^+$. Also, the definition of the local spectral function of Eq.~(\ref{spectral_function}) implies
\begin{eqnarray}
A(\omega) &=& -\frac{1}{\pi}\Im[G^{(0)}_R (\omega)] \nonumber \\
&=& \frac{1}{\mathcal{Z}^{(0)}} \sum_{n,m} \delta(\omega- \Delta E_{n\rightarrow m})|\bra{n} b\ket{m}|^2 e^{-\beta E_n} \nonumber \\
&& - \frac{1}{\mathcal{Z}^{(0)}} \sum_{n,m} \delta(\omega- \Delta E_{n\rightarrow m})|\bra{m} b\ket{n}|^2 e^{-\beta E_n} \;, \nonumber
\label{appendix_spectral_function}
\end{eqnarray}
as a consequence of the limit
\begin{eqnarray}
\lim_{\eta\rightarrow 0^+} \frac{1}{(\omega-\Delta E)+i\eta}  = -\pi\delta(\omega- \Delta E) \;.
\label{appendix_limmit_delta}
\end{eqnarray}
The explicit evaluation of the previous matrix elements yields
\begin{eqnarray}
A^{(0)}(\omega) &=& \frac{1}{\mathcal{Z}^{(0)}} \sum_n \delta (\omega -\Delta E_{n\rightarrow n+1}) (n+1) e^{-\beta E_n} \nonumber \\
&&-\frac{1}{\mathcal{Z}^{(0)}} \sum_n  \delta (\omega -\Delta E_{n-1\rightarrow n})\:n e^{-\beta E_n} \;,
\label{appendix_atomic_limit_local_spectral_function}
\end{eqnarray}
which shows the atomic limit spectral function as a collection delta functions at the consecutive transitions $\ket{n} \rightarrow \ket{n+1}$. This analysis makes explicit that the negative sign contributions of $A$ comes from the hole excitation branch.

\section{Correlations and free energy}\label{appendix:correlations_free_energy}

\indent In Subsec.~\ref{subsec:correlations} we presented how the free energy is obtained from the atom-atom correlation function according to Eq.~(\ref{delta_omega_correlation}). To demonstrate this relation, we parametrize the system's Hamiltonian in the form $H^{(\lambda)} = H^{(0)} +\lambda \mathcal{H} $ according to Eqs.~(\ref{hamiltonian_parameter}) and (\ref{hamiltonian_parameter_w}), interpolating the atomic limit $H^{(0)}$ and the original Hamiltonian  $H^{(1)}$ as $\lambda$ continuously varies between $0$ and $1$. \\
\indent The corresponding partition function is given by the trace
\begin{eqnarray}
\mathcal{Z}^{(\lambda)} = \Tr \left[e^{-\beta H^{(\lambda)}}\right] \;,
\label{appendix_partition_lambda}
\end{eqnarray}
from which we extract the free energy
\begin{eqnarray}
\Omega^{(\lambda)} = -\frac{1}{\beta}\ln \mathcal{Z}^{(\lambda)} \;.
\label{appendix_omega_lambda}
\end{eqnarray}
The derivative of $\Omega^{(\lambda)}$ with respect to $\lambda$ provides
\begin{eqnarray}
\frac{d\Omega^{(\lambda)}}{d\lambda} &=& -\frac{1}{\beta} \frac{\Tr \left [ (-\beta \mathcal{H})e^{-\beta H^{(\lambda)}}\right ]}{\Tr\left [ e^{-\beta H^{(\lambda)}}\right ]} \nonumber \\
&=&  \frac{\Tr \left [\mathcal{H} e^{-\beta H^{(\lambda)}}\right ]}{\Tr \left [ e^{-\beta H^{(\lambda)}} \right ]} = \langle \mathcal{H} \rangle_{\lambda} \;.
\label{appendix_omega_derivative}
\end{eqnarray}
On the other hand, the free energy variation follows immediately from the integration
\begin{eqnarray}
\int_0^{1}\frac{d\Omega^{(\lambda)}}{d\lambda} \: d\lambda = \Omega -\Omega^{(0)} = \Delta \Omega \;,
\label{appendix_delta_omega}
\end{eqnarray}
because $\Omega^{(1)} = \Omega$. By using the results of Eq.~(\ref{appendix_omega_derivative}) into Eq.~(\ref{appendix_delta_omega}), we get
\begin{eqnarray}
 \Delta \Omega = \int_0^{1}\ \langle \mathcal{H} \rangle_{\lambda} \: d\lambda \;.
\label{appendix_delta_omega_v}
\end{eqnarray}
Furthermore, the expression for the tunneling term $\mathcal{H}$ of Eq.~(\ref{hamiltonian_parameter_w}) yields
\begin{eqnarray}
\Delta \Omega &=& -J\sum_{\langle i,j \rangle}\int_0^{1} \langle {b}^{\dagger}_i{b}^{\vphantom{\dagger}}_j \rangle_{\lambda} \: d\lambda \nonumber \\
&=& -J\sum_{\langle i,j \rangle}\int_0^{1} \mathcal{C}^{(\lambda)}_{ij} \: d\lambda \;,
\label{appendix_delta_omega_correlation}
\end{eqnarray}
where
\begin{eqnarray}
\mathcal{C}^{(\lambda)}_{ij} = \langle {b}^{\dagger}_i{b}^{\vphantom{\dagger}}_j \rangle_{\lambda}
\label{appendix_correlation_lambda}
\end{eqnarray}
is the correlation function considering first neighbors $i$ and $j$, evaluated through a thermal average at a hopping $\lambda J$. For a homogeneous system the summation over all first neighbors yields $zN_s$, simplifying Eq.~(\ref{appendix_delta_omega_correlation}) to
\begin{eqnarray}
 \Delta \Omega = -zJN_s\int_0^{1} \mathcal{C}^{(\lambda)} \: d\lambda\;.
\label{appendix_delta_omega_correlation_homogeneous}
\end{eqnarray}

%

\end{document}